\documentclass[preprint,aps,floatfix,superscriptaddress,pre,showpacs]{revtex4}
\usepackage{amsmath, amsthm, amssymb, graphicx}
\input epsf.sty
\begin{document}

\def\g{\gamma}
\def\r{\rho}
\def\w{\omega}
\def\wo{\w_0}
\def\wp{\w_+}
\def\wm{\w_-}
\def\t{\tau}
\def\av#1{\langle#1\rangle}
\def\pf{P_{\rm F}}
\def\pr{P_{\rm R}}
\def\F#1{{\cal F}\left[#1\right]}

\title{Surface phase diagram of the three-dimensional kinetic Ising model in an oscillating magnetic field}

\author{Keith Tauscher and Michel Pleimling}
\affiliation{Department of Physics, Virginia Tech, Blacksburg, Virginia 24061-0435, USA}

\date{\today}

\begin{abstract}
We study the surface phase diagram of the three-dimensional kinetic Ising model below the 
equilibrium critical point subjected to a periodically 
oscillating magnetic field. Changing the surface interaction strength as well as the  
period of the external field, we obtain a non-equilibrium surface phase diagram that in parts strongly resembles the corresponding
equilibrium phase diagram, with an ordinary transition, an extraordinary transition and a surface transition.
These three lines meet at a special transition point. For weak surface couplings, however, the surface does not
order. These results are found to remain qualitatively unchanged when using different single-spin flip dynamics.
\end{abstract}

\maketitle

%\section{Introduction}
{\em Introduction.} The presence of surfaces can have a huge impact on local quantities close to a bulk critical point.
For example, in magnetic spin systems a surface can not
only change the local critical exponents, it can also yield a complicated surface diagram with different
surface transitions \cite{Bin83,Die86,Die97,Ple04}. At the ordinary transition the bulk alone is critical,
yielding surface critical exponents which have been computed systematically using field-theoretical 
techniques. In the presence of strongly enhanced surface couplings, the surface of a $d$-dimensional system
may order alone if the $(d-1)$-dimensional bulk system can order on its own. Lowering the temperature, this so-called
surface transition is followed by the extraordinary transition at which the bulk orders in 
presence of an already ordered surface. These three transition
lines meet at the special transition point at which both the surface and bulk correlation lengths
diverge. The presence of external fields can yield additional interesting phenomena as for example critical wetting.

However, much less is understood when non-equilibrium systems with surfaces are considered. Here we may distinguish
between systems that are prepared in a non-equilibrium initial state and allowed to relax toward equilibrium
and systems with a truly non-equilibrium steady state. In the former case, novel collective dynamical properties
emerge due to the simultaneous presence of a temporal and a spatial surface \cite{Rit95,Maj96,Ple04b,Mar12}.
Very few attempts have been made to investigate surface criticality of systems with non-equilibrium steady
states. These few studies focused on some absorbing phase transitions (as for example directed percolation)
\cite{Jan88,Ess96,How00,Fro01}
or on the kinetic Ising model subjected to an oscillating magnetic field \cite{Par12,Par13,Akt13}.

In this report we complement the investigation of the semi-infinite kinetic Ising model below the
equilibrium critical point in a periodically
oscillating magnetic field. In \cite{Par12} surface critical exponents were investigated numerically for both
the two- and three-dimensional systems in cases where the bulk and surface couplings have identical strengths \cite{Par13},
corresponding to the ordinary transition.
This study revealed that the surface exponents at the ordinary transition differ markedly from those obtained
from the corresponding equilibrium system. This is an interesting result, as it is well established that 
the bulk kinetic Ising model in a periodically
oscillating magnetic field belongs to the same universality class as the equilibrium Ising model \cite{Gin85,Kor00,Bue08,Par13b}.
Clearly, the presence of a surface at a non-equilibrium phase transition results in effects that are not yet
fully understood. 

The aim of the present study is two-fold. On the one hand, by changing the ratio between surface and bulk couplings we
want to elucidate numerically the surface phase diagram of the three-dimensional semi-infinite kinetic Ising model
(The reader should note that the phase diagram presented in Fig. 4b of Ref. \cite{Par12} is in fact the phase diagram
for $J_s = J_b$ kept fixed but with varying coupling strength between the surface layer and the underlying layer,
see the erratum \cite{Par13}). As we
will see, our study reveals a surface phase diagram very similar to the equilibrium model, with an ordinary transition,
a surface transition, an extraordinary transition, and a special transition point. This fully supports a recent study \cite{Akt13}
where in the framework of an effective field theory the existence of a special transition point was predicted. On the other
hand, we are also interested in understanding how a change of dynamics changes the steady-state properties of our system.
In an equilibrium system static properties are the same for every choice of the dynamics that fulfills detailed balance.
However, for a system with a non-equilibrium steady state, the choice of the dynamics can alter the properties of the system.
As we show in this report, going from Glauber to Metropolis dynamics does not change qualitatively the surface phase diagram and
has only small effects on the location of the phase transition lines. 

%\section{Model}
{\em Model.} 
The non-equilibrium phase transition encountered in magnetic systems below their equilibrium critical points
subjected to a periodically oscillating magnetic
field has attracted much interest, both theoretically \cite{Tom90,Cha99,Ach05,Sid98,Kor00,Par13b} and experimentally
\cite{Jia95,Rob08,Ber13}. This is a dynamic order-disorder phase transition where changing the period of the field allows 
the system to 
move from one phase to the other. When the period of the field is large, the spins are able to follow the magnetic field and
the magnetization averaged over one period is zero. This is the disordered phase. However, when the period of the field
is small, the system is not able to fully reverse its magnetization before the sign of the magnetic field changes
again. Consequently, the magnetization averaged over one period is no longer zero, which is the signature of the
ordered phase. Here 'small' and 'large' period is to be understood with respect to the metastable lifetime, which is the time needed
for the system to decay from a fully magnetized state in presence of a field pointing in the opposite direction.

Many insights regarding this non-equilibrium phase transition 
have been gained by studying Ising systems with periodic boundary conditions
in all space directions. In the following we consider the three-dimensional Ising model on a cubic lattice with free
boundary conditions in the $z$-direction and periodic boundary conditions in the $x$- and $y$-directions, thereby
introducing two surfaces in the $z$-direction. We allow for different interaction strengths in the surface layers as compared
to the interactions elsewhere in the system. The Hamiltonian is then
\begin{equation} \label{eq:h}
{\mathcal H} = - J_s \sum\limits_{\left[ {\bf x},{\bf y} \right]}
S_{\bf x} S_{\bf y} - J_b \sum\limits_{\left< {\bf x},{\bf y} \right>} S_{\bf x} S_{\bf y} 
- H(t) \sum\limits_{\bf x} S_{\bf x}~,
\end{equation}
where the first sum is exclusively over pairs of surface spins whereas the second term is over pairs of spins where at
least one spin is not in the surface layer. Here $S_{\bf x}= \pm 1$ is the Ising spin located at site ${\bf x}$.
The surface and bulk coupling constants are both ferromagnetic, i.e.
$J_s > 0$ and $J_b > 0$. The third term in (\ref{eq:h}) is due to the interaction of the spins with the time-dependent
external field. We use square wave fields with strength $H_0$ and half-period $t_{1/2}$. In this work we restrict
ourselves to values for the field amplitude and temperature used in previous
studies \cite{Kor00,Par12,Par13b}: $H_0 = 0.4 J_b$ and $T = 0.8 T_c$, where $T_c = 4.5115~J_b/k_B$ is the critical
temperature of the three-dimensional equilibrium Ising model. 

In all the simulations reported below we considered cubic systems with $L^3$ spins where $L$ ranges from
32 to 128.

Due to the presence of surfaces all quantities of interest depend on the distance to the surface. We therefore
study layer-dependent quantities, notably (i) the layer dependent order parameter
\begin{equation} \label{eq:OP}
Q(z) = \frac{1}{2 t_{1/2}} \oint m(z,t)\,dt~,
\end{equation}
i.e. the layer magnetization averaged over one period, where $m(z,t)$ is the magnetization of layer $z$ at time $t$,
(ii) the layer Binder cumulant
\begin{equation} \label{eq:U}
U(z) = 1 - \frac{\left< \left[ Q(z) \right]^4 \right>}{3 \left< \left[ Q(z) \right]^2 \right>^2}~,
\end{equation}
and (iii) the layer dependent scaled variance of the order parameter
\begin{equation} \label{eq:chi}
\chi(z) = L^{d-1} \left( \left< \left[ Q(z) \right]^2 \right> - \left< Q(z) \right>^2 \right)~.
\end{equation}
Here $\left< \cdots \right>$ indicates an average over many periods. The surface quantities are obtained for $z=1$ and
$z=L$, whereas we take as bulk quantities the quantities in the middle of the sample.

In order to better understand the effect of the chosen dynamics we
study two different single spin flip schemes, namely Glauber dynamics and Metropolis dynamics. 
After selecting a spin $S_{\bf x}$ at random, we compute the energy difference $\Delta E = {\mathcal H}(-S_{\bf x})
- {\mathcal H}(S_{\bf x})$ that would entail when flipping this spin. This change of configuration is then accepted
with the rate
\begin{equation} \label{eq:G}
w_G\left(S_{\bf x} \longrightarrow - S_{\bf x}\right) = \frac{\alpha_G}{1 + e^{\Delta E/k_B T}}
\end{equation}
for Glauber dynamics. For Metropolis dynamics a spin flip yielding a decrease of energy is always accepted, whereas
a spin flip yielding an increase of energy is accepted with rate
\begin{equation} \label{eq:M}
w_M\left(S_{\bf x} \longrightarrow - S_{\bf x}\right) = \alpha_M e^{-\Delta E/k_B T}~.
\end{equation}
$\alpha_G$ and $\alpha_M$ are constants that only fix the time scale. We make the
common choice $\alpha_G = \alpha_M=1$.

\begin{figure} [h]
\includegraphics[width=0.90\columnwidth]{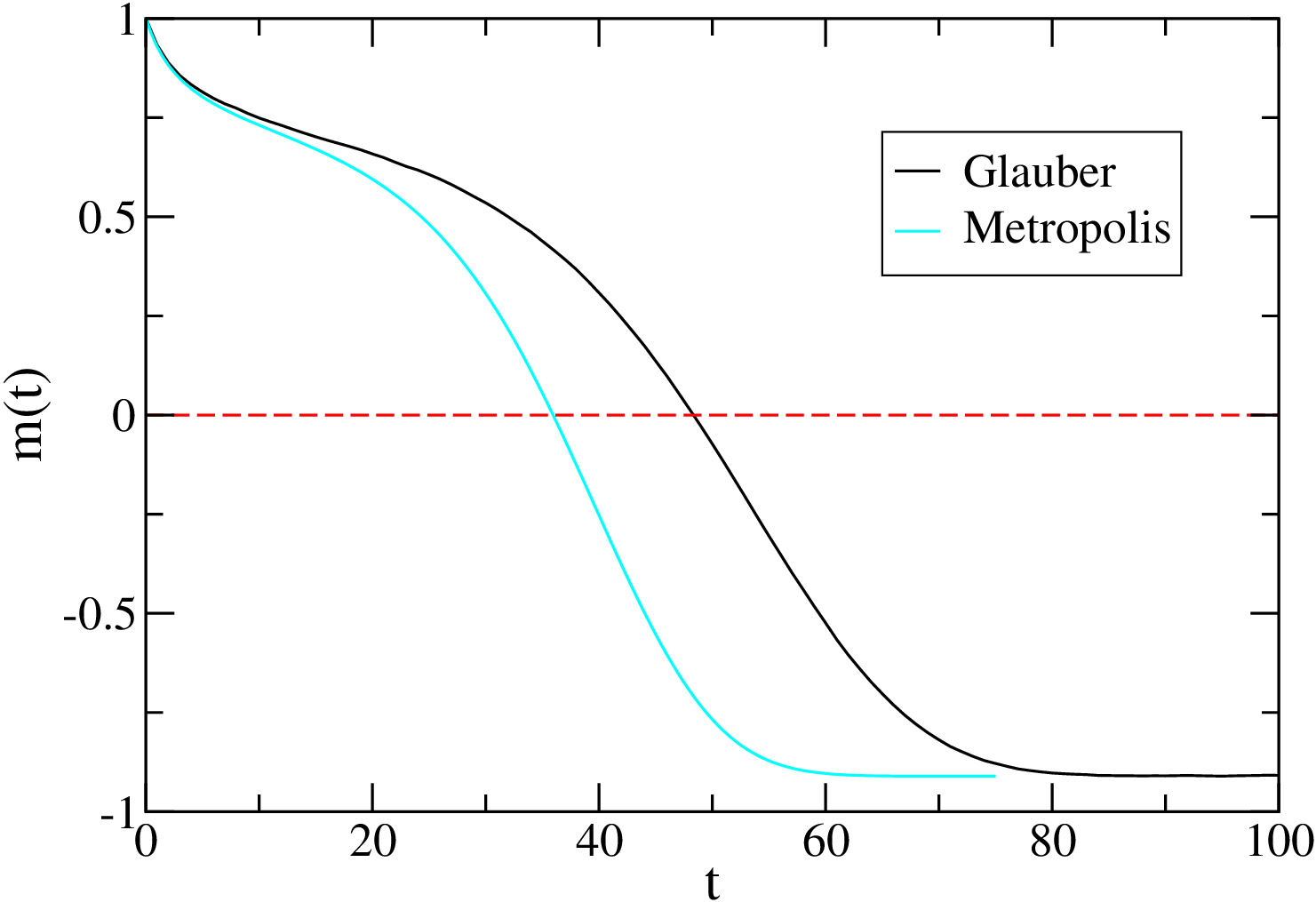}
\caption{\label{fig1} (Color online)
Determination of the metastable lifetime for both Glauber and Metropolis dynamics for $T = 0.8 T_c$
and $H_0 = 0.4 J$. The system is prepared in a fully positively magnetized state and a magnetic field
pointing in the negative direction is applied. The metastable lifetime is defined as the average time
the system needs to reach zero magnetization, as indicated by the dashed line. The data shown here, 
which result from averaging over 1000 independent runs, have been obtained
for systems with $L =128$, but the metastable lifetime is found to be independent of the system size.
}
\end{figure}

The important reference time for the following discussion is the metastable lifetime. This is the average time needed
for a fully magnetized sample to reach zero magnetization
when a magnetic field pointing in the opposite direction is applied. Fig. \ref{fig1} shows the time evolution
of such a fully magnetized sample for our parameters $T = 0.8 T_c$ and $H_0 = 0.4 J$. The intersection with
the zero magnetization line yields the metastable lifetimes $\tau = 47.05$ and $\tau = 35.95$ for Glauber and
Metropolis dynamics, respectively.

%%%%%%%%%%%%%%%%%%%%%%%%%%%%%%%%%%%%%%%%%%%FIG 1.%%%%%%%%%%%%%%%%%%%%%%%%%%%%%%%%%%%%%%%%%%%%%%%%%%%%%%
%%%%%%%%%%%%%%%%%%%%%%%%%%%%%%%%%%%%%%%%%%%FIG 1.%%%%%%%%%%%%%%%%%%%%%%%%%%%%%%%%%%%%%%%%%%%%%%%%%%%%%

%\section{Numerical results}
{\em Numerical results.}
In order to determine the surface phase diagram we consider surface couplings ranging from $J_s = J_b$
to $J_s = 2 J_b$. The data reported in the following result from averaging over typically $10^5$ periods of
the field after having reached the steady state. Error bars are of the order of the symbols used in
the figures.

An important quantity is the parameter $\Theta = \frac{t_{1/2}}{\tau}$ that describes the competition
between the oscillating magnetic field and the metastable state characterized by the lifetime $\tau$.
In our kinetic model this quantity takes over the role played by temperature in the equilibrium system: when
$\Theta$ increases (due to an increase of the half period $t_{1/2}$) a phase transition takes place between
a dynamically ordered and a dynamically disordered phase. When using Glauber dynamics, this transition takes
place at $\Theta_G = 1.285$ \cite{Par13b}. For Metropolis dynamics we locate the bulk transition point
at $\Theta_M = 1.257$.

%%%%%%%%%%%%%%%%%%%%%%%%%%%%%%%%%%%%%%%%%%%FIG 2.%%%%%%%%%%%%%%%%%%%%%%%%%%%%%%%%%%%%%%%%%%%%%%%%%%%%%%
\begin{figure} [h]
\includegraphics[width=0.90\columnwidth]{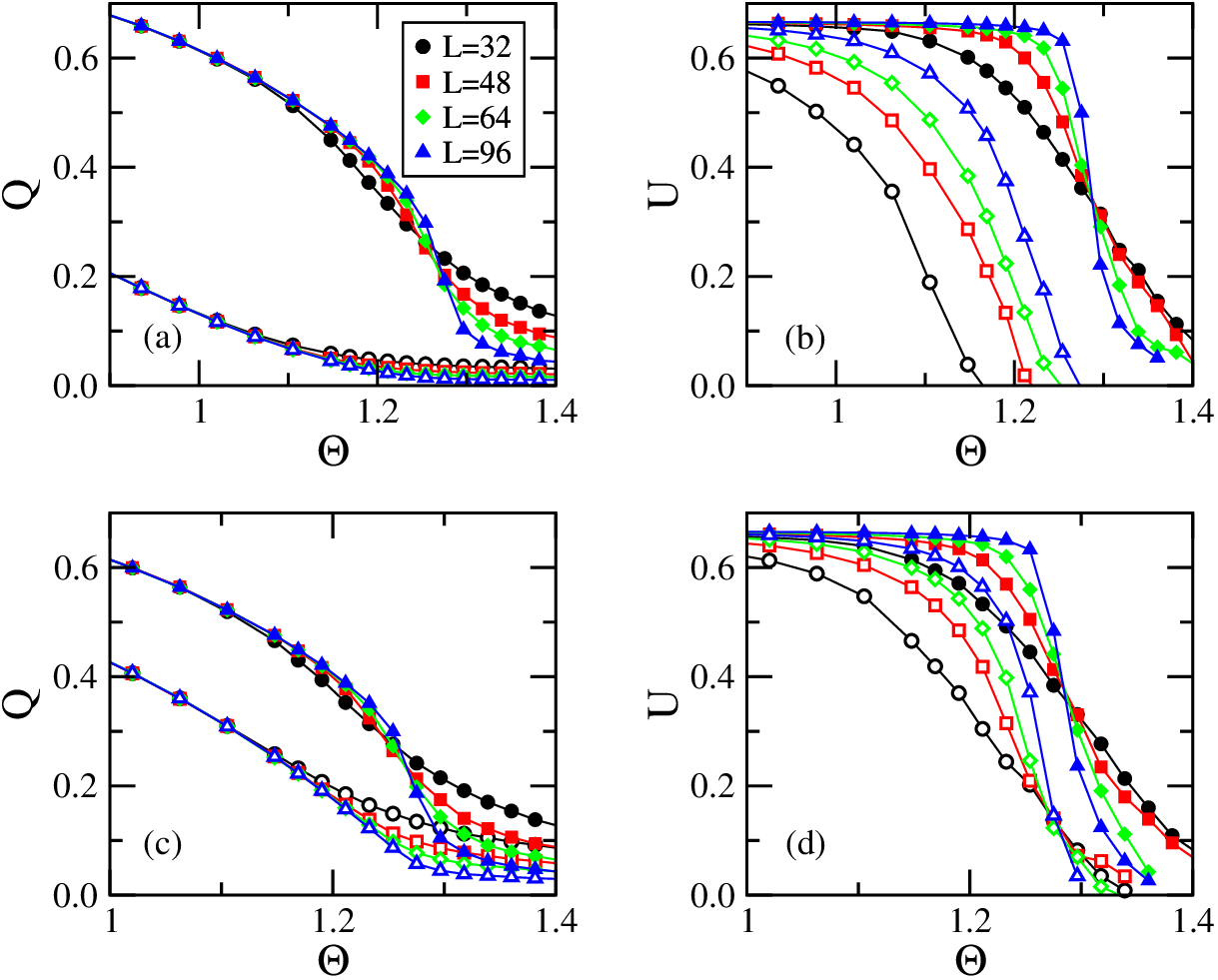}
\caption{\label{fig2} (Color online)
Surface (open symbols) and bulk (filled symbols) order parameter and Binder cumulant as a function of of the
competition parameter $\Theta$
for (a,b) $J_s = J_b$ and (c,d) $J_s = 1.25 J_b$, using Glauber dynamics. Data for different system sizes are shown.
}
\end{figure}
%%%%%%%%%%%%%%%%%%%%%%%%%%%%%%%%%%%%%%%%%%%FIG 2.%%%%%%%%%%%%%%%%%%%%%%%%%%%%%%%%%%%%%%%%%%%%%%%%%%%%%

%%%%%%%%%%%%%%%%%%%%%%%%%%%%%%%%%%%%%%%%%%%FIG 3.%%%%%%%%%%%%%%%%%%%%%%%%%%%%%%%%%%%%%%%%%%%%%%%%%%%%%%
\begin{figure} [h]
\includegraphics[width=0.90\columnwidth]{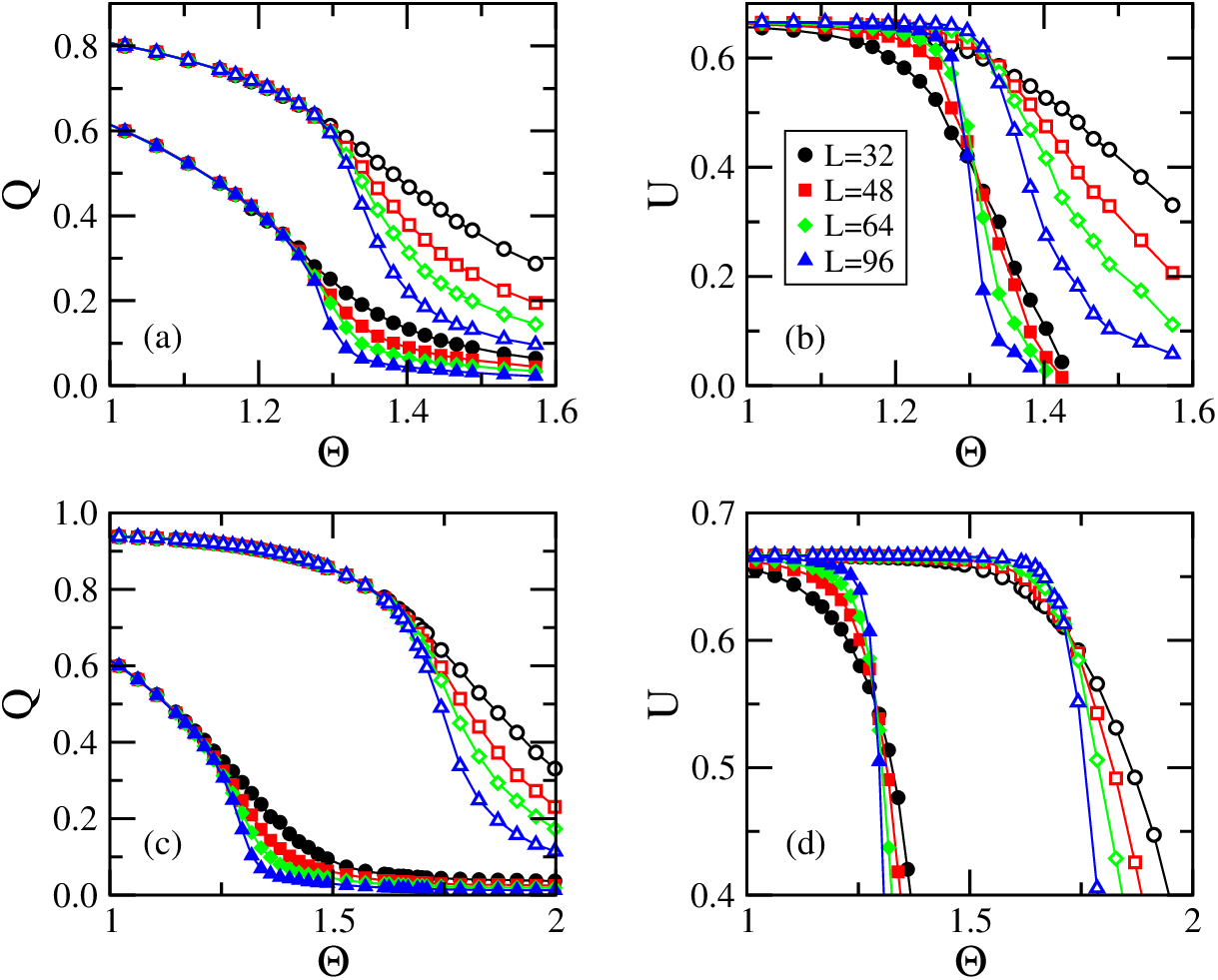}
\caption{\label{fig3} (Color online)
Surface (open symbols) and bulk (filled symbols) order parameter and Binder cumulant as a function of
the competition parameter $\Theta$
for (a,b) $J_s = 1.50 J_b$ and (c,d) $J_s = 1.75 J_b$, using Glauber dynamics. Data for different system sizes are shown. 
}
\end{figure}
%%%%%%%%%%%%%%%%%%%%%%%%%%%%%%%%%%%%%%%%%%%FIG 3.%%%%%%%%%%%%%%%%%%%%%%%%%%%%%%%%%%%%%%%%%%%%%%%%%%%%%

\begin{figure} [h]
\includegraphics[width=0.90\columnwidth]{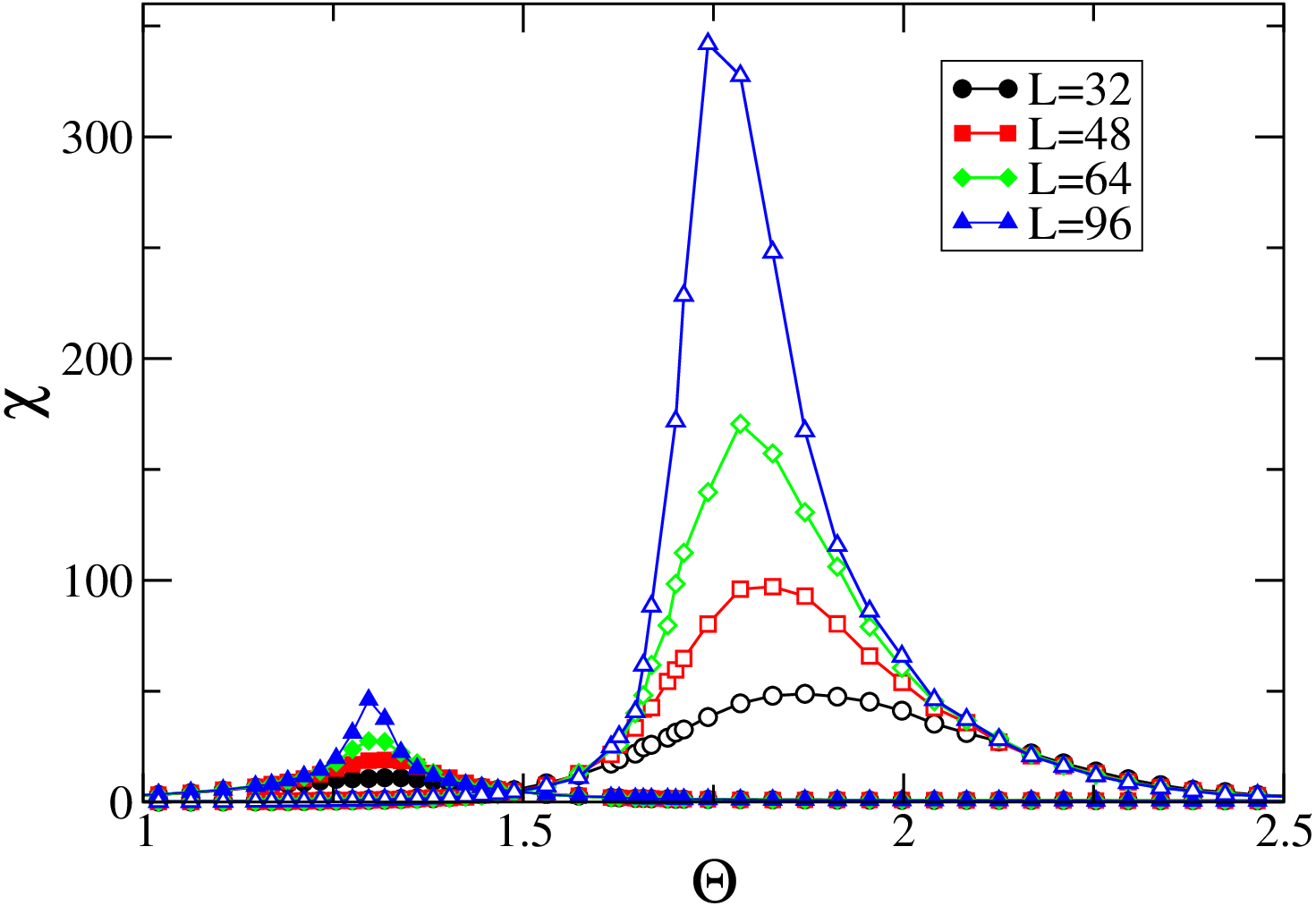}
\caption{\label{fig4} (Color online)
Scaled variance of the local order parameter
at the surface (open symbols) and in the middle of the system (filled symbols) for $J_s = 1.75
J_b$, using Glauber dynamics.
}
\end{figure}

As shown in Figs. \ref{fig2} and \ref{fig3} for the case of Glauber dynamics, different scenarios prevail depending
on the strength of the surface couplings. For weak surface couplings, with $J_s < 1.25 ~J_b$, the surface does not order
dynamically at the bulk transition point, as demonstrated by the 
surface Binder cumulant which does not exhibit a crossing of the lines obtained for different system sizes,
see Fig. \ref{fig2}b. Due to missing bonds surface spins can follow much easier the oscillating
magnetic field than bulk spins, and no surface ordering takes place at the bulk critical point. A partial dynamic
ordering is observed for lower values of $\Theta$, see Fig. \ref{fig2}a,
but this effect is not related to a phase transition. For surface 
couplings of intermediate strength, with $1.25~J_b \leq J_s < 1.45~J_b$, the surface orders at the bulk transition point, see
Figs. \ref{fig2}c and \ref{fig2}d. At this ordinary transition, the surface quantities display a singular behavior
governed by novel surface critical exponents, as discussed in \cite{Par12}. Finally, for strong surface couplings, 
see Fig.\ \ref{fig3}, the surface orders alone at values of $\Theta$ larger than the bulk transition
point, followed by the extraordinary transition where the bulk orders in presence of an already ordered surface.
These two different phase transitions are also clearly observed when studying the layer dependent variance of the 
order parameter, see Fig. \ref{fig4}.
Based on our data, we encounter the two distinct phase transitions for values of $J_s > 1.45~J_b$. This allows us
to locate the special transition point at $J_s \approx 1.45 J_b$.

%%%%%%%%%%%%%%%%%%%%%%%%%%%%%%%%%%%%%%%%%%%FIG 4.%%%%%%%%%%%%%%%%%%%%%%%%%%%%%%%%%%%%%%%%%%%%%%%%%%%%%%
%%%%%%%%%%%%%%%%%%%%%%%%%%%%%%%%%%%%%%%%%%%FIG 4.%%%%%%%%%%%%%%%%%%%%%%%%%%%%%%%%%%%%%%%%%%%%%%%%%%%%%

\begin{figure} [h]
\includegraphics[width=0.90\columnwidth]{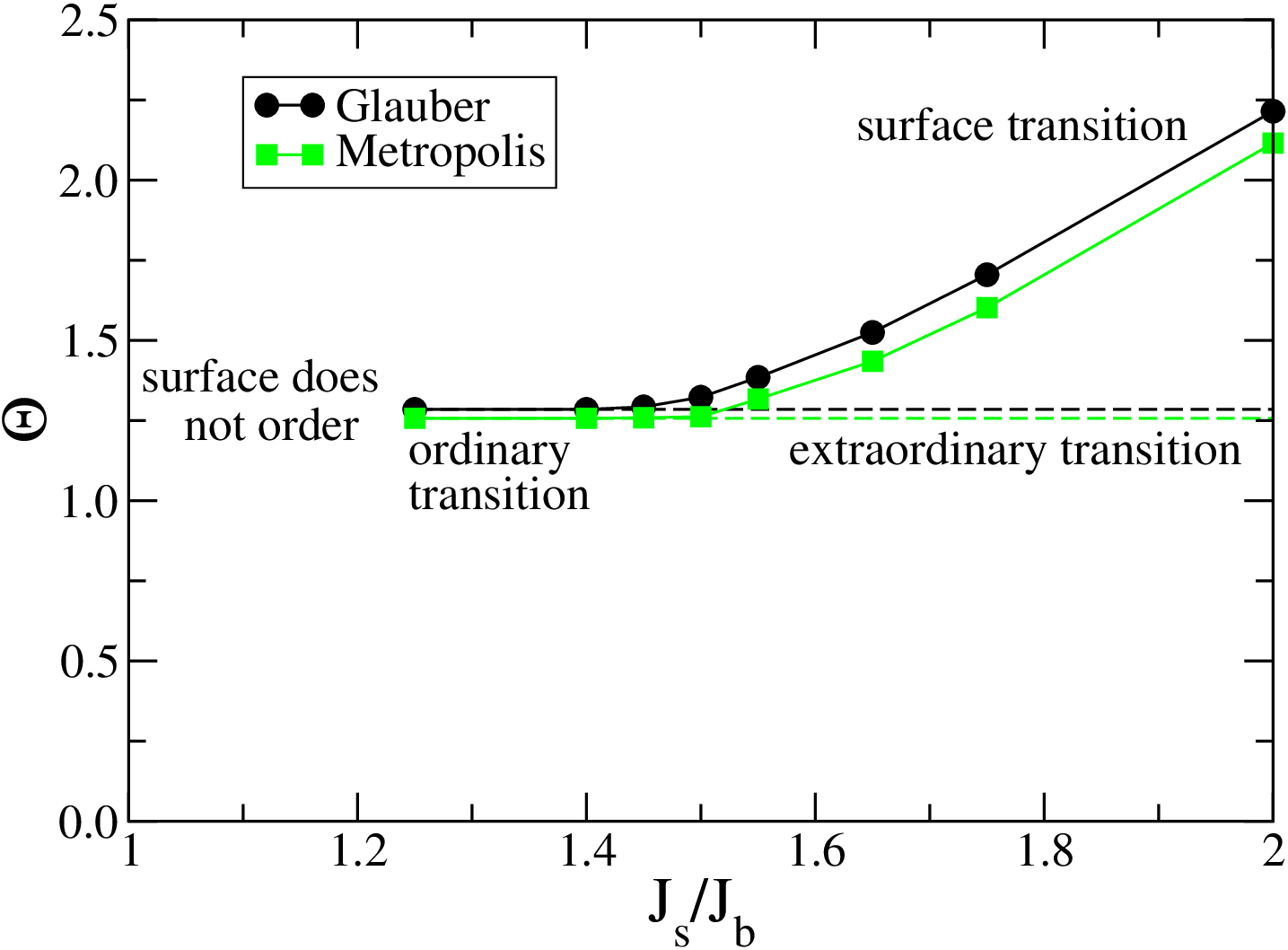}
\caption{\label{fig5} (Color online)
Surface phase diagram for both Glauber and Metropolis dynamics as a function of the coupling ratio
$J_s/J_b$ and the competition parameter $\Theta$ for $H_0 =0.4$ and $T=0.8~T_c$. The kinetic surface phase diagram resembles
the phase diagram of the equilibrium model, with an ordinary transition, a surface transition, and
an extraordinary transition. These three lines meet at the special transition point whose location for both
dynamics is at $J_s \approx 1.45 J_b$. For values of $J_s < 1.25 J_b$ the surface does not order
at the bulk critical point.
}
\end{figure}

Fig.\ \ref{fig5} shows the resulting dynamic surface phase diagram for the kinetic semi-infinite Ising model exposed to
a square wave field with field amplitude $H_0 =0.4$ at the temperature $T=0.8~T_c$. With the exception of the regime
of weak surface couplings, where no surface ordering takes place at the bulk transition point, this phase diagram
resembles very much the surface phase diagram observed for the semi-infinite equilibrium Ising model \cite{Ple04}.
We include in Fig.\ \ref{fig5} our results for both Glauber and Metropolis dynamics. The general features of
the surface diagram are independent of the chosen dynamics. Especially, for both schemes the special transition point
is found to be at $J_s \approx 1.45 J_b$. The only quantitative differences are given by small shifts of the phase transition
lines, mainly due to the fact that for Metropolis dynamics the bulk transition takes place at a sightly smaller
value of $\Theta$ than for Glauber dynamics.

%%%%%%%%%%%%%%%%%%%%%%%%%%%%%%%%%%%%%%%%%%%FIG 5.%%%%%%%%%%%%%%%%%%%%%%%%%%%%%%%%%%%%%%%%%%%%%%%%%%%%%%
%%%%%%%%%%%%%%%%%%%%%%%%%%%%%%%%%%%%%%%%%%%FIG 5.%%%%%%%%%%%%%%%%%%%%%%%%%%%%%%%%%%%%%%%%%%%%%%%%%%%%%

%\section{Conclusion}
{\em Conclusion.}
Our study reveals that the non-equilibrium surface phase diagram of the ordered three-dimensional
kinetic Ising model in a periodically oscillating field exhibits all the phase transitions encountered
in the corresponding equilibrium models. The existence of a special transition point is in agreement
with the effective field calculations presented in \cite{Akt13}. A new feature in the
non-equilibrium system is the absence of surface ordering at the bulk transition point 
for weak surface couplings, which is due to the physical mechanism underlying the ordering process.

\begin{acknowledgments}
This work is supported by the US National
Science Foundation through grant DMR-1205309. We thank H. Park for his early contributions to this
project and U. C. T\"{a}uber for a critical reading of the manuscript.
\end{acknowledgments}


\begin{thebibliography}{99}
\bibitem{Bin83} K. Binder, in Phase Transitions and Critical Phenomena, edited by C. Domb and
J.L. Lebowitz (Academic Press, London, 1983), vol. 8.
\bibitem{Die86} H. W. Diehl, in Phase Transitions and Critical Phenomena, edited by C. Domb and
J.L. Lebowitz (Academic Press, London, 1986), vol. 10.
\bibitem{Die97} H. W. Diehl, Int. J. Mod. Phys. B {\bf 11}, 3503 (1997).
\bibitem{Ple04} M. Pleimling, J. Phys. A: Math. Gen. {\bf 37}, R79 (2004).
\bibitem{Rit95} U. Ritschel and P. Czerner, Phys. Rev. Lett. {\bf 75}, 3882 (1995).
\bibitem{Maj96} S. N. Majumdar and A. M. Sengupta, Phys. Rev. Lett. {\bf 76}, 2394 (1996).
\bibitem{Ple04b} M. Pleimling, Phys. Rev. B {\bf 70}, 104401 (2004).
\bibitem{Mar12} M. Marcuzzi, A. Gambassi, and M. Pleimling, EPL {\bf 100}, 46004 (2012).
\bibitem{Jan88} H.-K. Janssen, B. Schaub, and B. Schmittmann, 
Z. Phys. B {\bf 71}, 377 (1988).
\bibitem{Ess96} J. W. Essam, A. J. Guttmann, I. Jensen, and D. TanlaKishani,
J. Phys. A: Math. Gen. {\bf 29}, 1619 (1996).
\bibitem{How00} M. Howard, P. Fr\"{o}jdh, and K. B. Lauritsen, 
Phys. Rev. E {\bf 61}, 167 (2000).
\bibitem{Fro01} P. Fr\"{o}jdh, M. Howard, and K. B. Lauritsen, 
Int. J. Mod. Phys. B {\bf 15}, 1761 (2001).
\bibitem{Par12} H. Park and M. Pleimling, Phys. Rev. Lett. {\bf 109}, 175703 (2012).
\bibitem{Par13} H. Park and M. Pleimling, Phys. Rev. Lett. {\bf 110}, 239903(E) (2013).
\bibitem{Akt13} B. O. Akta\c{s}, \"{U}. Ak\i nc\i, and H. Polat, arXiv:1302.2727.
\bibitem{Gin85} G. Grinstein, C. Jayaprakash, and Y. He, Phys. Rev. Lett. {\bf 55}, 2527 (1985).
\bibitem{Kor00} G. Korniss, C. J. White, P. A. Rikvold, and M. A. Novotny, Phys. Rev. E {\bf 63}, 016120 (2000).
\bibitem{Bue08} G. M. Buend\'{i}a and P. A. Rikvold, Phys. Rev. E {\bf 78}, 051108 (2008).
\bibitem{Par13b} H. Park and M. Pleimling, Phys. Rev. E {\bf 87}, 032145 (2013).
\bibitem{Tom90} T. Tom\'{e} and M. J. de Oliveira, Phys. Rev. A {\bf 41}, 4251 (1990).
\bibitem{Cha99} B. Chakrabarti and M. Acharyya, Rev. Mod. Phys. {\bf 71}, 847 (1999).
\bibitem{Ach05} M. Acharyya, Int. J. Mod. Phys. C {\bf 16}, 1631 (2005).
\bibitem{Sid98} S. W. Sides, P. A. Rikvold, and M. A. Novotny, Phys. Rev. Lett. {\bf 81}, 834 (1998).
\bibitem{Jia95} Q. Jiang, H.-N. Yang, and G.-C. Wang, Phys. Rev. B {\bf 52}, 14911 (1995).
\bibitem{Rob08} D. T. Robb, Y. H. Xu, O. Hellwig, J. McCord, A. Berger, M. A. Novotny, and P. A. Rikvold,
Phys. Rev. B {\bf 78}, 134422 (2008).
\bibitem{Ber13} A. Berger, O. Idigoras, and P. Vavassori, Phys. Rev. Lett. {\bf 111}, 190602 (2013).
\end{thebibliography}
\end{document}